\documentstyle[pre,preprint,aps,psfig]{revtex}

\begin{document}

\draft

\title{Field and intensity correlation in random media}

\author{P. Sebbah$^1$, R. Pnini$^2$ and A.Z. Genack$^3$}
\address{
$^1$Laboratoire de Physique de la Mati\`ere Condens\'ee/CNRS UMR
6622
\\Universit\'e de Nice-Sophia Antipolis, Parc Valrose, 06108
Nice Cedex 02, France
\\ $^2$Technion-Israel Institute of
Technology, 32000 Haifa, Israel
\\ $^3$Department of Physics, Queens
College of the City University of New York, Flushing, New York
11367, USA}

\date{\today}

\maketitle

\begin{abstract}
We have obtained the spectral and spatial field correlation
functions, $C_E(\Delta\omega)$ and  $C_E(\Delta x)$,
respectively, from the measurement of the microwave field
spectrum at a series of points along a line on the output of a
random dielectric medium. $C_E(\Delta\omega)$ and $C_E(\Delta x)$
are shown to be the Fourier transforms, respectively, of the time
of flight distribution, obtained from pulsed measurements, and of
the specific intensity. Unlike $C_E(\Delta\omega)$, the imaginary
part of $C_E(\Delta x)$ is shown to vanish as a result of the
isotropy of the correlation function in the output plane. The
complex square of the field correlation function gives the short
range or $C_1$ contribution to the intensity correlation function
$C$. Longer-range contributions to the intensity correlation
function are obtained directly by subtracting $C_1$ from $C$ and
are in good agreement with theory.
\end{abstract}


The random phasing of multiply-scattered partial waves in
disordered media leads to intensity fluctuations whose
short-range spatial correlation is determined by the square of
the field correlation function. This field-factorization term is
the leading or $C_1$ contribution to the cumulant intensity
correlation function $C = \left<\delta I \delta I'\right> \sim
C_1 = |\left<EE'^*\right>|^2$. Here $\delta I$ is the fluctuation
of the intensity $I$ from its ensemble average value, $\delta I =
I - \left<I\right>$, $E$ is the electromagnetic field, and
$\left<\ldots\right>$ represents the ensemble average. This
approximation gives the well-known Gaussian statistics of the
field \cite{Goodman,Chabanov97} and the negative exponential
distribution of the intensity
\cite{GoodmanDainty,Shapiro,Garcia89}. The intensity correlation
length $\delta x$ sets the scale of the speckle spots, while the
correlation frequency $\delta\nu$ represents the inverse of the
spread in transit time between the source and the point of
observation. Recent advances in speckle statistics have shown
however that fluctuations in intensity
\cite{Garcia89,stretched,OLMarin,RossumNieuwenhuizen,Kogan95,Falko95},
total transmission
\cite{RossumNieuwenhuizen,Stephen,Feng4,Berkovits,Ad90,Ad92,Ad94,PRLMarin}
and conductance
\cite{Laibowitz,LeeStone,Altshuler,AltshulerAltshuler,Prigodin93}
are enhanced beyond the predictions of Gaussian statistics, as a
result of long and infinite-range intensity correlations, the
$C_2$ and $C_3$ contributions to the intensity correlation
function, respectively.

In this paper, we present observations of the field correlation
function $\left<EE'^*\right>$ in the spectral and spatial domains
in measurements of the microwave field transmitted through random
media. We relate these to their corresponding Fourier transforms,
the time of flight distribution of photons and the specific
intensity, which gives the angular distribution of the intensity
in the far field. Taking the complex square of the field
correlation function gives $C_1$ contribution and makes it
possible to consider the difference $C-C_1$, which includes only
terms beyond the field factorization approximation, which are
responsible for enhanced fluctuations of transmission quantities.
Since higher order contributions than $C_2$ are small in these
samples, $C-C_1$ essentially equals $C_2$. In the past $C_2$ was
found from measurements of the correlation function of total
transmission \cite{Ad90,Ad92} or from fitting the measured
intensity correlation function using an assumed functional form
for $C_1$ and $C_2$ \cite{GGW}. Here $C_2$ is directly separated
from the intensity correlation function by subtracting $C_1$ from
$C$. Theoretical expressions for $C_1$ and $C_2$ in the frequency
domain including absorption and boundary conditions are in good
agreement with experiments.

\bigskip
In order to eliminate instrumental distortions of field and
intensity spectra, we normalize the field by the square root of
the average intensity and the intensity by its average value at
each frequency and position. In addition, measurements are carried
out in a frequency range in which the change in scattering
parameters is small. This allows us to obtain the functional form
of the relative degree of correlation. We will henceforth denote
the normalized intensity $I/\left<I\right>$ by $I$ and the
normalized field $E/\left<I\right>^{1/2}$  by $E$. The cumulant
correlation function of the normalized intensity will now be
denoted by $C$, which is the sum of three terms
\cite{Mello,MAS,Feng4,Berkovits}:
\begin{equation}
C = C_1 + C_2 + C_3 \ .
\label{C1+C2+C3}
\end{equation}
The $C_1$ term  is unity for frequency shift $\Delta \nu = 0$ and
displacement $\Delta x = 0$. The spectral field correlation
function for a plane wave incident upon a slab or upon a
quasi-one-dimensional sample is given by
\cite{Drake,GenackSheng,PniniS,GGPS}:
\begin{equation}
C_E(\Delta\nu) = \left<E(\nu)E^*(\nu+\Delta\nu)\right> =
\frac{\sinh q_0a}{\sinh q_0L'}\times\frac{\sinh \alpha L'}{\sinh
\alpha a} \ ,
 \label{CE}
\end{equation}
where $L$ is the sample length, $L' = L + 2z_b$ is the effective
length corrected for internal reflection, $z_b=2\ell(1+R)/3(1-R)$
is the extrapolation length \cite{Ad89,ZhuPineWeitz,GenackLi}, $R$
is the reflection coefficient at the boundary averaged over
internal incident angles, $\ell$ is the transport mean free path,
$\alpha=1/L_a$ is the inverse of the diffusive absorption length
$L_a=\sqrt{D\tau_a}$, $D$ is the diffusion constant, $\tau_a$ is
the absorption time, $a=5\ell/3$ is the randomization distance
from the plane in which the intensity inside the medium
extrapolates to zero, $q_0=\gamma_+-i\gamma_-$,
$\gamma^2_\pm={1\over2}(\sqrt{\alpha^4+\beta^4}\pm\alpha^2)$ and
$\beta=\sqrt{2\pi\Delta\nu/D}$. The $C_2$ term is of order $1/g$
at $\Delta \nu = 0$ and $\Delta x = 0$. Far from the localization
transition, the expression for the $C_2$ contribution including
absorption is given by \cite{PniniS91,Kogan92}:
\begin{equation}
C_{2}(\Delta \nu)=\frac{4}{gL} \frac{\sqrt{\alpha ^{4}+\beta
^{4}}}{\beta^{4}} \frac{\gamma _{+}\sinh \left( 2\gamma
_{+}L\right) - \gamma_{-}\sin \left( 2\gamma _{-}L\right) -\alpha
\sinh \left( 2\alpha L\right) }{\cosh \left( 2\gamma _{+}L\right)
-\cos \left( 2\gamma _{-}L\right) }\, \label{C2} \ ,
\end{equation}
where $g=Ak_{0}^{2}\ell /3\pi L$ is the ensemble averaged
dimensionless conductance for a tube of length $L$ and cross
section $A$. This expression differs from Eq.~(14) in
\cite{PniniS91} by a factor of 2 to take into account the
one-channel-in, one-channel-out measurement described bellow.
Thus,
\begin{equation}
C_2(\Delta\nu=0)=\frac{\sinh 2\alpha L -2\alpha L(2-\cosh2\alpha
L)} {4g\alpha L\sinh^2(\alpha L)} \ . \label{gprime}
\end{equation}
In the absence of absorption, $C_2(\Delta \nu) = \frac{4}{3g}$.
Equations~(\ref{C2}) and (\ref{gprime}) do not include internal
reflection. The complete expression for $C_2$ including internal
reflection is detailed in the Appendix~A. The $C_3$ contribution
is responsible for universal conductance fluctuations measured in
electronic systems
\cite{Laibowitz,LeeStone,Altshuler,AltshulerAltshuler}. It has
been recently observed in optics in time correlation experiments
\cite{Scheffold98}. This contribution is of order $1/g^2$ and is
more strongly suppressed by absorption than the $C_2$ term
\cite{RossumNieuwenhuizen} and will therefore not be considered
here.

\bigskip
The system studied here is composed of randomly positioned
${1\over2}$-inch polystyrene spheres at a volume filling fraction
of 0.52 contained within a one meter long, 7.6 cm diameter copper
tube. Wire antennas are used as the emitter and detector at the
input and output surfaces of the sample. A Hewlett Packard 8722C
vector network analyzer is used to measure the microwave field,
giving its amplitude and phase. Measurements are made in the Ku
and K bands with frequency steps of 625 kHz and displacements of
1 mm along a 4 cm line running symmetrically about the center of
the output surface. After the spectrum is taken at a given
position, the detector is translated by 1 mm. Measurements are
made for two positions of the input antenna separated by 3 cm.
Once spectra are taken at each point on the line, the copper tube
is rotated briefly to create a new configuration. Measurements
are made in 200 sample realizations and spectral and spatial
correlation functions are computed from these data.

The real and imaginary parts of the field correlation function
with frequency shift are shown in Fig.~\ref{fig1}. Averaging has
been performed over sample configuration, position of the source
and detector and over frequency between 16.8 and 17.8 GHz, where
transport parameters are known to vary slowly \cite{GenackLi}.
The real part of the correlation function falls quadratically,
whereas the imaginary part rises linearly with frequency shift,
for small shifts \cite{vanT}. From a three parameters fit using
Eq.~(\ref{CE}) and the value of the reflection coefficient
$R=0.13$ measured in \cite{GenackLi}, we find the diffusion
constant $D=3.3\times 10^{10} \text {cm}^2$/s, the absorption
length $L_a=33.3$ cm, the penetration depth $a=18.3$ cm, which in
turn give the extrapolation length $z_b=9.5$ cm and the mean free
path $\ell=11.0$ cm.

We have also carried out measurements of pulsed propagation in the
same sample. The response to a 1 ns pulse shaped by a pulse
forming network and mixed to a local oscillator at 19 GHz is
measured using high bandwidth B\&H Electronics amplifiers. The
results are collected using a digital sampling oscilloscope. The
complex square is taken and averaged over 4096 configurations and
shown in Fig.~\ref{fig2}. Because the incident pulse is much
shorter than the typical traversal time through the medium, and
encompasses a bandwidth much greater than the correlation
frequency, this measurement gives the time of flight distribution
of photons propagating through the sample at a carrier frequency
of 19 GHz. The line through the data is the Fourier transform of
the field correlation function which is averaged over
configuration and frequencies between 18.5 and 19.5 GHz. The
agreement between these sets of data shows that the field
correlation function and the time of flight distribution are
Fourier transform pairs in agreement with the demonstration in
Appendix~B. This result is analogous to the demonstration that
the intensity distribution on the output surface due to point
excitation (the point spread function) and the intensity
correlation function with identical shifts in the incident and
scattered wave vector \cite{Feng4,Freund88} are Fourier transform
pairs \cite{Li94}. In both cases the variables of the incident
and outgoing waves are shifted by the same amount. This result
holds even in the presence of significant long-range intensity
correlation. In previous optical studies, \cite{Drake}, only the
intensity was measured. Its correlation function with frequency
was shown to be the square of the Fourier transform of the time
of flight distribution. In that case, long-range correlation was
negligible and the measured correlation function was essentially
equal to $C_1$.

The real and imaginary parts of the field correlation function
with displacement $C_E(\Delta x)=\left<E(x,\nu)E^*(x+\Delta
x,\nu)\right>$ averaged over sample configuration, frequency
$\nu$ between 18 and 18.5 GHz, and position $x$, are shown in
Fig.~\ref{fig3}. The imaginary part of the field correlation
function with displacement is small, in contrast to the field
correlation function with frequency (see Fig.~\ref{fig1}). As
shown below, this is a consequence of the isotropy of the random
field. Such isotropy may be expected at the output of a
quasi-one-dimensional sample, far from the boundary, or on length
scales smaller than the sample thickness at the output of a
random slab. By the translational invariance of the field
correlation function with displacement, $C_E(\Delta x) =
\left<E(x)E^*(x+\Delta x)\right> = \left<E(x- \Delta
x)E^*(x)\right> = C_E^*(-\Delta x)$. By the isotropy of $C_E
(\Delta x)$ with regard to the direction of the displacement, we
can exchange $-\Delta x$ and $\Delta x$. This gives $C_E(\Delta
x) = C_E^*(\Delta x)$. Since $C_E(\Delta x)$ equals its complex
conjugate, its imaginary part vanishes. In contrast, the
imaginary part of the field correlation function with frequency
shift $C_E(\Delta\omega)$ does not vanish because the change of
$\Delta\omega$ to $-\Delta\omega$ also requires that the complex
conjugate be taken.

In transmission, the angular distribution of the scattered
intensity is predicted to be \cite{Freund}:
\begin{equation}
I(\theta)=\Delta cos\theta + cos^2\theta \label{I(theta)} \ ,
\label{Morse}
\end{equation}
where $\Delta=z_b/l$ and the scattering angle $\theta$ is measured
from the normal to a reference plane $\Sigma$ close to the output
surface. Following Ref.~\cite{Freund}, the 2-D Fourier transform
of this expression is the 2-D spatial correlation function of the
field over $\Sigma$. When taken along a line, this reduces to
\begin{equation}
C_E(\Delta x)=\frac{1}{(1+2\Delta)}\left[\Delta \frac{\sin(k_0
\Delta x)}{k_0 \Delta x} + \frac{J_1(k_0 \Delta x)}{k_0 \Delta
x}\right]  \ ,
\label{Freund}
\end{equation}
where $k_0$ is the free space k-vector. This expression gives a
good fit over 30 points to the data (Fig.~\ref{fig3}) for
$\Delta=0.73$ and $k_0=3.6$ cm$^{-1}$, in good agreement with the
calculated value at 18 GHz of $3.8$ cm$^{-1}$. This confirms that
the scattered intensity angular distribution $I(\theta)$ and the
spatial correlation function $C_E(\Delta x)$ are Fourier transforms.

\bigskip
Squaring the field gives the intensity and allows us to compute
the intensity correlation function and to compare it to the
modulus square of the field correlation function. In the
frequency domain, we compare the measured intensity correlation
function $C(\Delta\nu)$ to $C_1+C_2$ given in Eqs.~(\ref{CE}) and
(\ref{CompleteC2}), assuming $C_3$ is negligible and using the
value of $L_a$, $D$, $a$ and $z_b$ found from the fit to $C_E$
without additional adjustable parameters (Fig.~\ref{fig4}). The
value $C(\Delta\nu=0)=1.13$ is the variance of the normalized
intensity. The difference of this from unity is a measure of
mesoscopic corrections to correlation. The $C-C_1$ contribution
normalized by its value at $\Delta\nu=0$ is shown in
Fig.~\ref{fig5}. The comparison to Eq.~(\ref{CompleteC2}), which
includes $z_b$, and to Eq.~(\ref{C2}) which does not, shows the
effect of internal reflection.

\bigskip
In conclusion, we have measured the field correlation function
with frequency and displacement and shown that they are the
Fourier transforms of the time of flight distribution and the
specific intensity, respectively. Comparison with theoretical
expressions gives the diffusion coefficient, diffusive absorption
length, extrapolation length and mean free path. The measurement
of both the field and intensity correlation functions makes
possible a direct separation between short and long-range
intensity correlation.

\bigskip
\acknowledgments We are indebted to the late Narciso Garcia for
valuable suggestions, for support and encouragement. We thank
Prof. Boris Shapiro and O. Legrand for helpful discussions and
Marin Stoytchev for technical contributions. This work was
supported by the National Science Foundation under Grant Nos. DMR
9973959 and INT9512975, a PSC-CUNY grant, the United
States-Israel Binational Science Foundation (BSF) and the
Groupements de Recherche POAN and PRIMA.

\bigskip
\appendix
\section{}

We give the complete expression for the long range contribution
$C_2$ including absorption and internal reflection, for quasi-one
dimensional geometry. We follow the approach of
\cite{RossumNieuwenhuizenPLA} in which the correlation function
was written in the form:
\begin{equation}
C_2 (\Delta\nu)=2\frac{N_a+N_b + N_\alpha}{gBL} \ ,
\label{CompleteC2}
\end{equation}
with an overall factor of 2 to account for the one channel-in, one
channel-out experimental setup. We obtain the following
expressions, which differ for $B$ and $N_\alpha$ from
\cite{RossumNieuwenhuizenPLA}:

\bigskip
\begin{mathletters}
\begin{eqnarray}
B&=&\left[ 1 + 2z_b^2( {\gamma_+^2}-{\gamma_-^2}) +
z_b^4(\gamma_+^2+\gamma_-^2)^2\right]
\left[\cosh (2 \gamma_+ L) -\cos(2 \gamma_- L)\right]  \nonumber \\
&+& 4z_b^2 (\gamma_+^2 +\gamma_-^2)\left[\cosh (2 \gamma_+ L) +\cos(2 \gamma_- L)\right]  \nonumber \\
&+&\left\{4\gamma_+z_b\left[ 1 + z_b^2( {\gamma_+^2} +
{\gamma_-^2})\right] \sinh(2\gamma_+L) \right\}+
\left\{4\gamma_-z_b\left[ 1 - z_b^2( {\gamma_+^2} +
{\gamma_-^2})\right] \sin (2\gamma_-L)\right\} \ , \\
N_a&=&\frac{1}{2\gamma_+(\gamma_+^2-\alpha^2)} \nonumber \\
&\times&\{ \left[ (2\gamma_+^2 -\alpha^2) + z_b^2(8\gamma_+^4 +
2\gamma_+^2\gamma_-^2 -\gamma_-^2\alpha^2
-3\gamma_+^2\alpha^2+\alpha^4) +
z_b^4(\gamma_+^2 +\gamma_-^2)(2\gamma_+^4-2\gamma_+^2\alpha^2 +\alpha^4)\right]  \nonumber \\
&&\sinh (2 \gamma_+ L)  \nonumber \\
&+&\left[4z_b\gamma_+ (3\gamma_+^2-\alpha^2) +4z_b^3 \gamma_+
(3\gamma_+^4+ \gamma_+^2\gamma_-^2 -2\gamma_+^2\alpha^2 +\alpha^4)\right] \sinh^2 (\gamma_+ L)\} \ , \\
N_b&=&\frac{1}{2\gamma_-(\gamma_-^2+\alpha^2)} \nonumber \\
&\times&\{\left[ -(2\gamma_-^2 +\alpha^2)
 + z_b^2(8\gamma_-^4 +2\gamma_+^2\gamma_-^2 +\gamma_+^2\alpha^2+3\gamma_-^2\alpha^2+\alpha^4)
- z_b^4(\gamma_+^2 +\gamma_-^2)(2\gamma_-^4+2\gamma_-^2\alpha^2 +\alpha^4)\right]  \nonumber \\
&&\sin (2 \gamma_- L) \nonumber \\
&+&\left[4z_b\gamma_- (3\gamma_-^2+\alpha^2) -4z_b^3 \gamma_-
(3\gamma_-^4+ \gamma_+^2\gamma_-^2 +2\gamma_-^2\alpha^2 +\alpha^4)\right] \sin^2 (\gamma_-L)\} \ , \\
N_{\alpha}&=&\frac{\gamma_+^2 +\gamma_-^2}{2\alpha(\gamma_+^2-\alpha^2)(\gamma_-^2+\alpha^2)}  \nonumber \\
&\times&\{ \left[ -\alpha^2  -z_b^2(6\gamma_+^2\gamma_-^2
+5\gamma_+^2\alpha^2-5\gamma_-^2\alpha^2+\alpha^4) + z_b^4\alpha^2
(-2\gamma_+^2\gamma_-^2 -\gamma_+^2\alpha^2 +\gamma_-^2\alpha^2)\right] \sinh (2 \alpha L)  \nonumber \\
&+&\left[4z_b\alpha^{-1}(-\gamma_+^2\gamma_-^2 -\gamma_+^2
\alpha^2 +\gamma_-^2\alpha^2-\alpha^4) + 4z_b^3
\alpha(-3\gamma_+^2\gamma_-^2 -2\gamma_+^2\alpha^2
+2\gamma_-^2\alpha^2)\right] \sinh^2 (\alpha L)\} \ .
\end{eqnarray}
\end{mathletters}

\section{}

We show that the time of flight distribution and the field
correlation function with frequency shift are Fourier transform
pairs. The response to a wave packet $f(t)$ is $E(t)= \int d\nu
\exp[-i2\pi\nu t] F(\nu) E(\nu)$, where $F(\nu)$ is the Fourier
transform of $f(t)$ and $E(\nu)$ is the field measured at a given
point at frequency $\nu$. From this expression, the ensemble
average of the temporal intensity variation of the transmitted
pulse may be written as,
\begin{equation}
\langle E^2(t)\rangle =\int d\nu_1 d\nu_2\exp[-i2\pi(\nu_1
-\nu_2)t] \langle E(\nu_1) E^\ast (\nu_2)\rangle F(\nu_1) F^\ast
(\nu_2) \ . \label{averagedintensity}
\end{equation}
If $f(t)$ is a short pulse with a bandwidth much wider than the
correlation frequency $\delta\nu$, $F(\nu_1)$ and $F(\nu_2)$ are
nearly identical as long as $\nu_2-\nu_1$ is much smaller than
the pulse bandwidth. With the change of variables
$\nu_1=\nu+\Delta\nu /2$ and $\nu_2=\nu-\Delta\nu/2$,
$F(\nu_1)F^\ast(\nu_2)\approx|F(\nu)|^2$ and
Eq.~(\ref{averagedintensity}) becomes
\begin{equation}
\langle E^2(t)\rangle =\int d\nu d\Delta\nu \exp[-i2\pi\Delta\nu
t] \langle E(\nu + \Delta\nu/2) E^\ast (\nu -\Delta\nu/2 )\rangle
|F(\nu)|^2 \ .
\end{equation}
Within a frequency range in which the variation in the averaged
intensity $\langle I(\nu)\rangle$ is small, we can write $\langle
E(\nu + \Delta\nu /2) E^\ast (\nu -\Delta\nu /2 )\rangle =
\langle I(\nu)\rangle C_E(\nu,\Delta\nu)$, where
$C_E(\nu,\Delta\nu)$ is the normalized field correlation function
around the frequency
 $\nu$. In a frequency range in which the change in
scattering parameters is small, $C_E(\nu,\Delta\nu)  =
C_E(\Delta\nu)$ and we find
\begin{eqnarray}
\langle E^2(t)\rangle &=& \int d\nu\langle I(\nu)|F(\nu)|^2\rangle
\int d\Delta\nu \exp[-i2\pi\Delta\nu~t] C_E(\Delta\nu)
 \\
&=& \int_0^\infty dt \langle I(t)\rangle\int d\Delta\nu
\exp[-i2\pi\Delta\nu~t] C_E(\Delta\nu) \ , \nonumber
\end{eqnarray}
where the last step follows from Parseval's theorem. Consequently,
the time of flight distribution ${\cal I}(t)=\langle
E^2(t)\rangle/\int dt \langle I(t)\rangle$ and the field
correlation function are Fourier transform pairs:
\begin{equation}
{\cal I}(t)=\int d\Delta\nu~C_E(\Delta\nu)\exp[-i2\pi\Delta\nu
~t] \ . \label{TF}
\end{equation}


\begin{figure}

\vspace{1.cm} \caption{Real (crosses) and imaginary (dots) part of
field correlation function with frequency shift. Eq.~(\ref{CE})
has been used to fit (solid line) the real part. The fit to the
imaginary part is not shown, but it necessarily follows.}

\label{fig1}
\end{figure}

\begin{figure}

\vspace{1.cm} \caption{Time of flight distribution for carrier
frequency 19 GHz. The continuous line is the Fourier Transform of
the field correlation function (Eq.~(\ref{TF})) in the range 18.5
to 19.5 GHz}

\label{fig2}
\end{figure}

\begin{figure}

\vspace{1.cm} \caption{Real (circles) and imaginary (dots) part of
field correlation function with displacement. Eq.~(\ref{Freund})
has been used to fit (solid line) the real part. The imaginary
part is predicted to vanish.}

\label{fig3}
\end{figure}

\begin{figure}

\vspace{1.cm} \caption{Intensity correlation function with
frequency shift (circles). The solid line is the theoretical
expression for $C_1+C_2$ using the values of $L_a$, $D$, $a$ and
$z_b$ found from the fit of $C_E$. The $C_1$ and $C_2$
contributions are represented by the dotted and dashed lines,
respectively.}

\label{fig4}
\end{figure}

\begin{figure}

\vspace{1.cm} \caption{Long range contribution to intensity
correlation function $C-C_1$ normalized to its value at
$\Delta\nu=0$. The solid line corresponds to
Eq.~(\ref{CompleteC2}) which includes internal reflection. The
dashed line is given by Eq.~(\ref{C2}), which does not include
internal reflection.}

\label{fig5}
\end{figure}

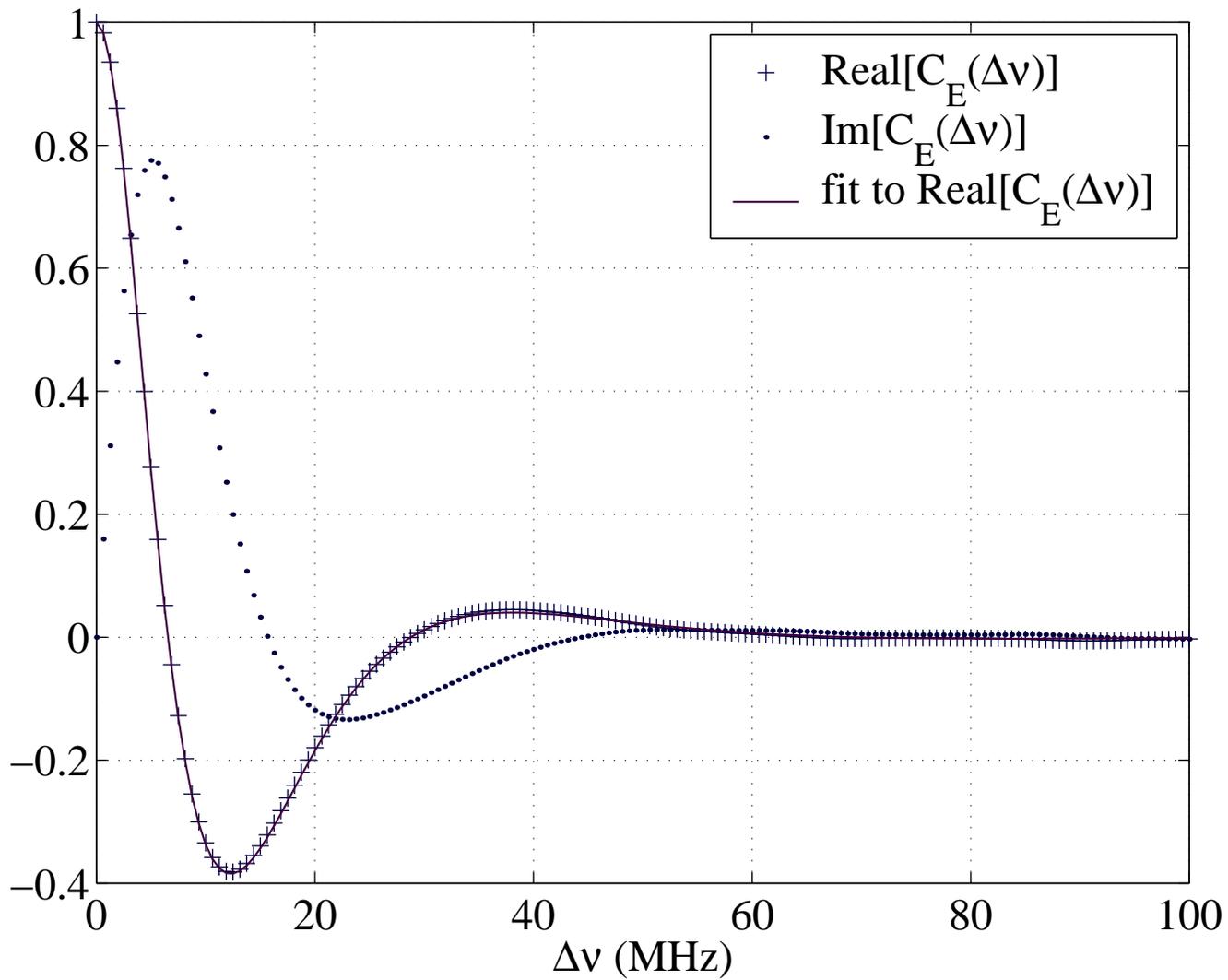
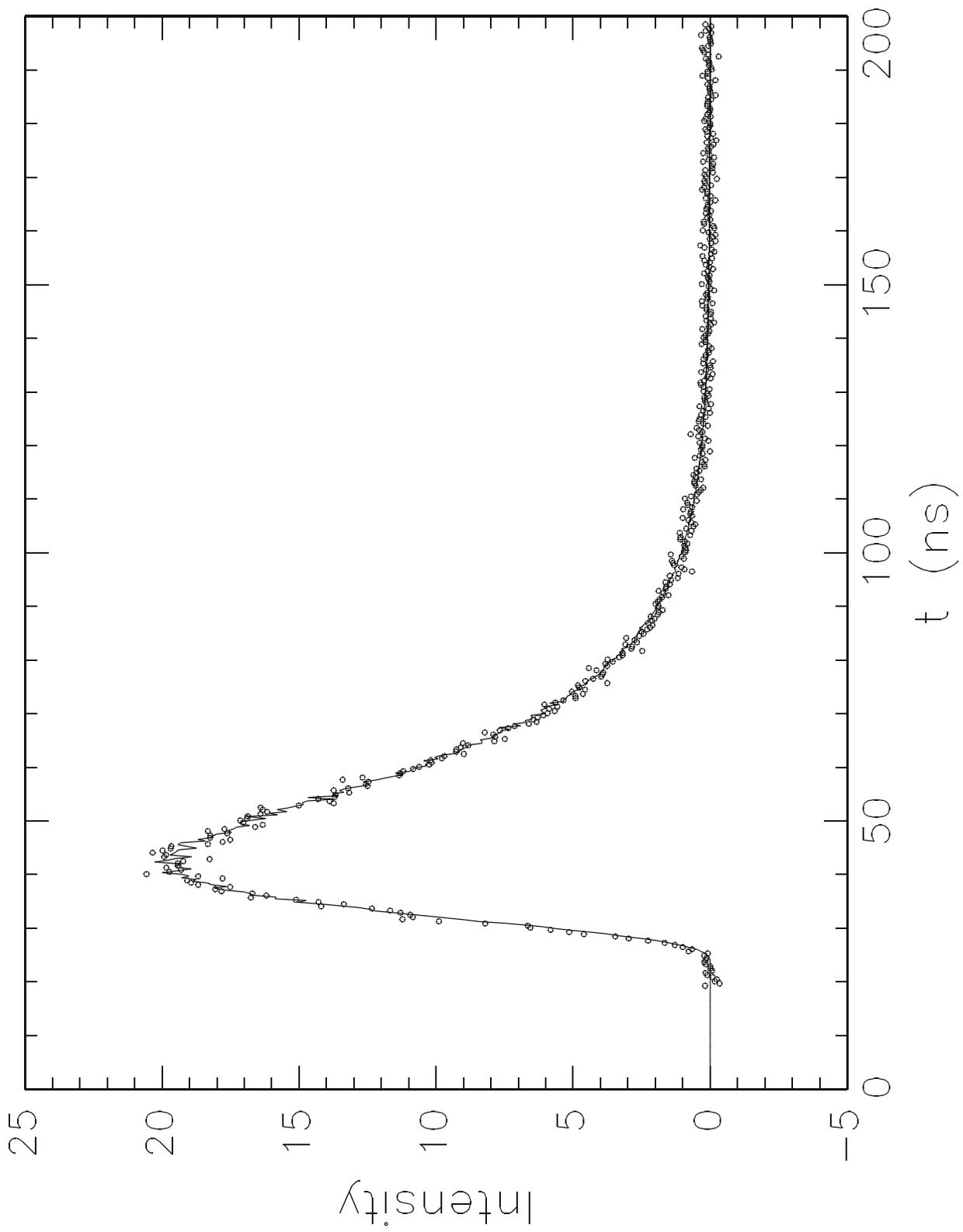
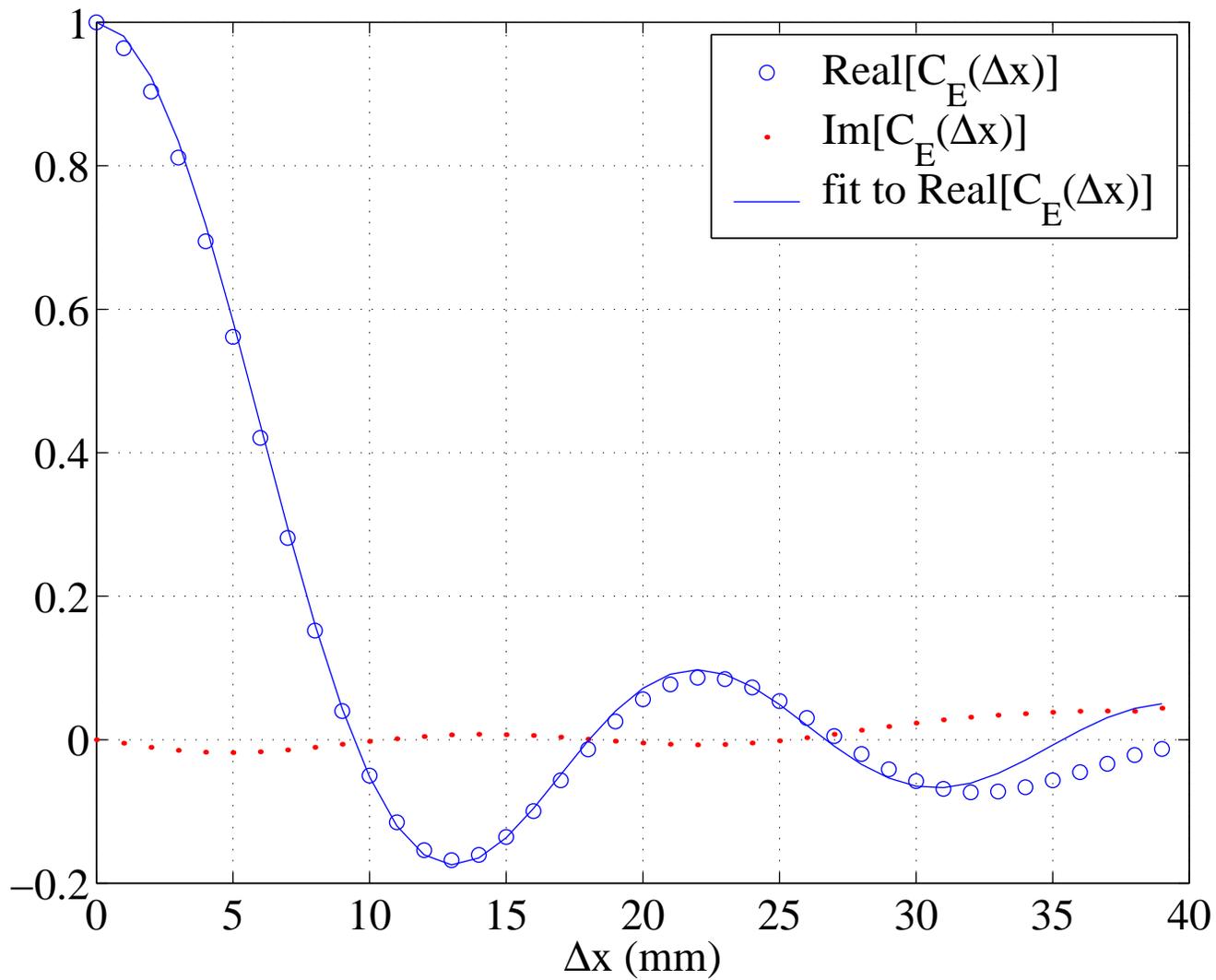
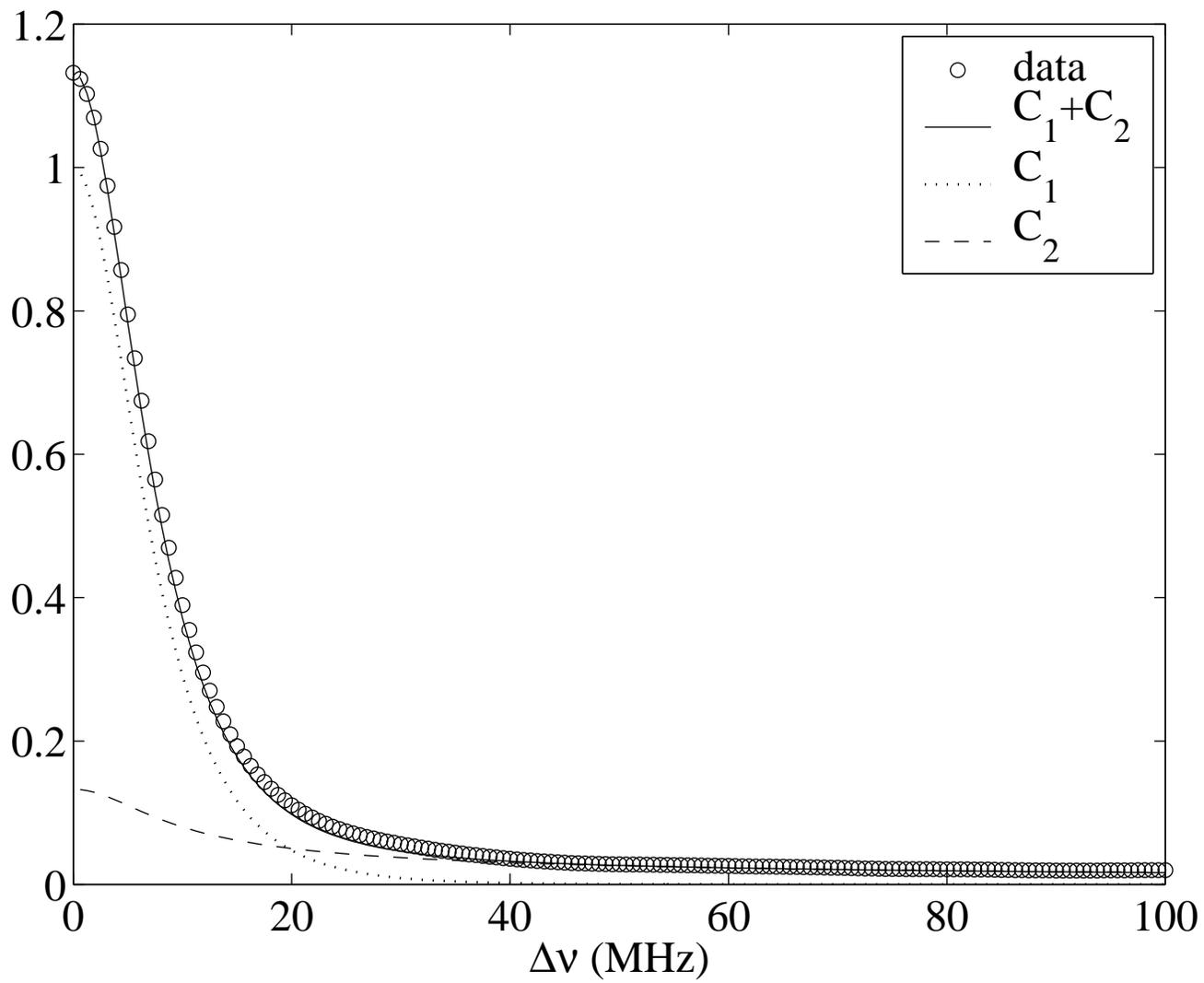
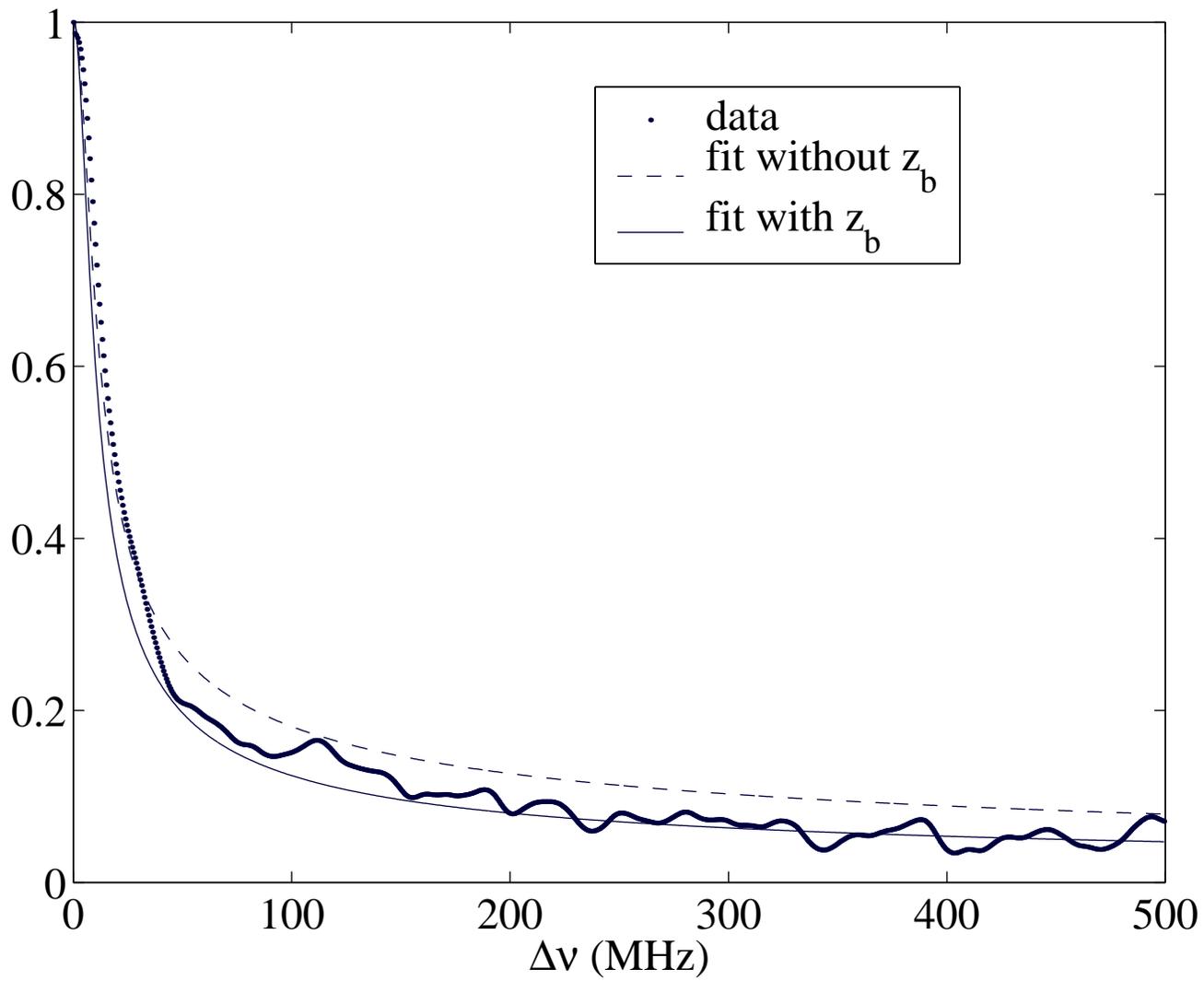
\end{document}